\documentclass[12pt]{article}
\usepackage{a4wide}
\usepackage{amssymb}
\usepackage{graphicx}
\begin{document}
{\renewcommand{\thefootnote}{\fnsymbol{footnote}}
\hfill  IGPG--06/8--1\\
\medskip
\hfill gr--qc/0609034\\
\medskip
\begin{center}
{\LARGE Loop quantum cosmology and inhomogeneities\footnote{Published
in Gen.\ Rel.\ Grav.\ (submitted: May 9, 2006)}}\\
\vspace{1.5em}
Martin Bojowald\footnote{e-mail address: {\tt bojowald@gravity.psu.edu}}
\\
\vspace{0.5em}
Institute for Gravitational Physics and Geometry,\\
The Pennsylvania State
University,\\
104 Davey Lab, University Park, PA 16802, USA\\
\vspace{1.5em}
\end{center}
}

\setcounter{footnote}{0}

\newcommand{\vertex}{v}
\newcommand{\vertexalt}{v'}
\newcommand{\case}[2]{{\textstyle \frac{#1}{#2}}}
\newcommand{\lP}{l_{\mathrm P}}

\newcommand{\md}{{\mathrm{d}}}
\newcommand{\tr}{\mathop{\mathrm{tr}}}
\newcommand{\sgn}{\mathop{\mathrm{sgn}}}

\newcommand*{\R}{{\mathbb R}}
\newcommand*{\N}{{\mathbb N}}
\newcommand*{\Z}{{\mathbb Z}}
\newcommand*{\Q}{{\mathbb Q}}
\newcommand*{\C}{{\mathbb C}}

\begin{abstract}
 Inhomogeneities are introduced in loop quantum cosmology using
 regular lattice states, with a kinematical arena similar to that in
 homogeneous models considered earlier.  The framework is intended to
 encapsulate crucial features of background independent quantizations
 in a setting accessible to explicit calculations of perturbations on
 a cosmological background. It is used here only for qualitative
 insights but can be extended with further more detailed input.  One
 can thus see how several parameters occuring in homogeneous models
 appear from an inhomogeneous point of view. Their physical roles in
 several cases then become much clearer, often making previously
 unnatural choices of values look more natural by providing
 alternative physical roles. This also illustrates general properties
 of symmetry reduction at the quantum level and the roles played by
 inhomogeneities. Moreover, the constructions suggest a picture for
 gravitons and other metric modes as collective excitations in a
 discrete theory, and lead to the possibility of quantum gravity
 corrections in large universes.
\end{abstract}

\section{Introduction}

Loop quantum gravity \cite{Rov,ALRev,ThomasRev} is a candidate for a
non-perturbative and background independent quantization of general
relativity which by now has uncovered several crucial features
relevant at small length scales. Most importantly, it unambiguously
leads to discreteness of spatial geometry at the Planck scale
\cite{AreaVol,Area,Vol2} which has played crucial roles in physical
pictures studied so far, such as cosmological and black hole
singularities \cite{Sing,BHInt,SphSymmSing}. These applications,
however, also require a significant amount of dynamical aspects which
remain poorly understood and quite complex in a full setting. Many
applications have therefore been formulated in symmetric models of
loop quantum cosmology \cite{SymmRed,LivRev}, which is a reduction
preserving crucial properties such as the very discreteness of the
full theory but at the same time making calculations explicitly
treatable. Results concerning new behavior at small scales then indeed
follow at least qualitatively as direct consequences of such basic
properties related to discreteness.

As homogeneous models in this framework are becoming studied in more
and more detail, initially designed structures have to be refined. In
particular, inhomogeneities are the crucial ingredient for most
remaining open issues. While background independence is conceptually
important and crucial for basic properties exploited in loop
applications, it also makes more detailed derivations of physical
properties complicated. A relation between quantum and classical
geometry is needed for nearly every aspect considered in quantum
gravity such as the singularity issue (where a relation to classical
geometry is required in identifying states where a classical
singularity would occur) or corrections to cosmological structure
formation. While full quantum geometry as it occurs at a fundamental
level provides many advanced techniques, a direct use usually comes
with too much baggage for a given application.  Additional
specializations to pick the right regime are often required. Examples
are symmetric states in considerations of singularities, or boundary
conditions for black hole entropy calculations
\cite{ABCK:LoopEntro,IHEntro} and in recent attempts to make contact
with low energy scattering \cite{BoundaryGraviton}. One is thus often
led to introduce special states corresponding to the physical
situation at hand, which allows one to make a detailed relation to
classical properties.  These states, in turn, provide useful
background structures.  As has been demonstrated by now in many
investigations, characteristic properties of the background
independent framework remain intact even when such additional
structures are introduced to extract and focus on physical regimes of
interest. Such a narrowing-down is a technical step to perform
explicit calculations or to find suitable approximation schemes,
rather than a conceptual part of the underlying theory. Given the
complexity as well as incompleteness of the full framework, a
systematic derivation of dynamics of such sectors from the full theory
is currently difficult, but many qualitative and semi-quantitative
aspects can be studied in carefully designed models.  If such
restrictions are not made, immense technical difficulties have to be
faced. For cosmological perturbation theory for instance the averaging
problem \cite{AveragingProblem}, which is even difficult in classical
gravity, arises while difficulties relevant for the singularity
problem have been described in \cite{DegFull}.  The aim is certainly
to arrive, at some point, at a complete relation to, if not a
derivation from, a fully background independent theory. But such a
derivation as well as the construction of the full theory itself will
be much easier if likely properties are already known and have been
evaluated in simpler situations.\footnote{Eddington's observation
``\ldots that it is indeed helpful in the quest for knowledge if we
understand the nature of the knowledge we are looking for''
\cite{Eddingtonknowledge} also applies in this case.}

An inclusion of inhomogeneous degrees of freedom in applications is
currently under construction, and already available for midisuperspace
models \cite{SphSymm,SphSymmVol,SphSymmHam} and in perturbative form
\cite{InhomEvolve}. From these developments it is becoming clear that
properties of inhomogeneous degrees of freedom have a bearing also on
strictly homogeneous models.  While the overall findings in the
homogeneous case are confirmed, the inhomogeneous analysis suggests
changes in constructions which one would not necessarily have
considered from within those models. A relation between homogeneous
and inhomogeneous models can in particular influence what is seen as
natural or unnatural in the constructions, and which corrections
should be expected to dominate.

We will discuss here basic properties of quantum variables used in
loop models, and draw four conclusions regarding parameters of
isotropic models when they are viewed as arising from inhomogeneous
ones. The setting we mainly have in mind is that of perturbative
inhomogeneities and thus refer particularly to the behavior of models
when they describe larger universes rather than regimes close to
classical singularities. Most of our observations, however, will be
general and can be checked non-perturbatively in midi-superspace
models, although they are justified by the specific formulation
introduced here only in regimes of perturbative inhomogeneities.  We
will also discuss technical details of symmetry reductions, suggest a
new picture for how metric modes on a background geometry such as
gravitons arise effectively through collective quantum excitations,
and indicate the possibility of quantum corrections to the universe
evolution at large scales.

\section{Isotropic loop quantum cosmology}

A classical isotropic model is fully described by the scale factor
$a(t)$ as a solution to the Friedmann equation once the matter content
is specified. In Ashtekar variables \cite{AshVar,AshVarReell} on which
loop quantum gravity is based the scale factor and its time derivative
are expressed through a densitized triad component $\tilde{p}$ with
$|\tilde{p}|=a^2$, determining an isotropic densitized triad
$E^a_i=\tilde{p}\delta^a_i$, and a connection variable
$\tilde{c}=\gamma\dot{a}$, determining an isotropic connection
$A_a^i=\tilde{c}\delta_a^i$ \cite{IsoCosmo} (with the Barbero--Immirzi
parameter $\gamma$ \cite{AshVarReell,Immirzi} relating the Ashtekar
connection to extrinsic curvature). The densitized triad encodes
spatial geometry as it is related to the spatial metric by
$E^a_iE^b_i=q^{ab}\det q$. In particular, the spatial volume of a
region $R$ is $V=\int_R\md^3x\sqrt{|\det E|}$ which for an isotropic
triad reduces to $V=V_0|\tilde{p}|^{3/2}=:|p|^{3/2}$ where the
coordinate volume $V_0$ is absorbed in $p$. The re-scaled variable
$p$, and similarly $c=V_0^{1/3}\tilde{c}$, is then independent of
coordinates in contrast to $\tilde{p}$ or the commonly used scale
factor $a=\sqrt{|\tilde{p}|}$ \cite{Bohr}. It does, however, depend on
the coordinate size $V_0$ of the region integrated over which can be
fixed to be the total space in a closed model but needs to be chosen
in open models where integrations have to be restricted to finite
regions. The relation of basic variables to metric or connection
components then depends on which choice is being made
here. Nevertheless, $V_0$ will not appear explicitly when one
restricts oneself to the variables $c$ and $p$ only, for instance the
Poisson relations
\[
 \{\tilde{c},\tilde{p}\}=\frac{8\pi\gamma G}{3V_0}
\]
become independent of $V_0$ after rescaling:
\[
 \{c,p\}=\frac{8\pi\gamma G}{3}\,.
\]

Following the loop quantization, these basic objects are represented
on a Hilbert space with an orthonormal basis
$\{|\mu\rangle\}_{\mu\in\R}$ of states given as functions of the
connection component by $\langle c|\mu\rangle = e^{i\mu c/2}$
\cite{Bohr}. Exponentials $e^{i\mu'c/2}$, analogous to holonomies of
the full theory, then act by multiplication,
\begin{equation}
 \widehat{e^{i\mu'c/2}}|\mu\rangle = |\mu+\mu'\rangle
\end{equation}
and the triad component $p$ by a derivative,
\begin{equation}
 \hat{p}|\mu\rangle = \frac{1}{6}\gamma\lP^2\mu|\mu\rangle\,.
\end{equation}
This demonstrates how basic properties of the full theory are
preserved: Only exponentials of $c$ are represented while it is not
possible to obtain an operator for an individual component $c$
directly, and the triad operator $\hat{p}$ has a discrete spectrum
since its eigenstates are normalizable. From $\hat{p}$ one directly
obtains the volume operator $\hat{V}=|\hat{p}|^{3/2}$.

Using the basic operators one then has to construct more complicated
ones relevant for dynamics. For matter Hamiltonians, inverse triad
components such as $|\det E|^{-1/2}$ are required which cannot simply
be obtained by taking an inverse operator of $\hat{p}$ because this
inverse does not exist: $\hat{p}$ has a discrete spectrum containing
zero. Nonetheless, one can construct operators corresponding to the
classical inverse using Poisson identities such as \cite{QSDV}
\begin{equation} \label{bracket}
 \frac{1}{|p|}= \left|\frac{2i}{\mu_0 l}e^{i\mu_0 c/2} \{e^{-i\mu_0
c/2},|p|^l\}\right|^{\frac{1}{1-l}}
\end{equation}
for $0<l<1$ and some $\mu_0$. On the right hand side, only a positive
power of $p$ is required which is easily available upon
quantization. The Poisson bracket will then become a commutator,
resulting in a well-defined and for isotropic models even bounded
operator \cite{InvScale}. The main effect is that inverse powers of
$p$ occuring in classical equations are replaced by regular functions,
which can be read off from eigenvalues of the operators of the form
\begin{equation}\label{inveigen}
  \left(\frac{1}{|p|}\right)_{\mu}\propto \left||\mu+\mu_0|^l-
|\mu-\mu_0|^l\right)^{1/(1-l)}
\end{equation}
and do not blow up at $p=0$. For instance, in a scalar matter
Hamiltonian, $|p|^{-3/2}$ occurs in the kinetic term which is replaced
by a regular function $d(p)=d_{l}(p/p_*)$ where $l$ is the ambiguity
parameter above and $p_*=\frac{1}{6}\gamma j\mu_0\lP^2$ with a second
parameter $j$ arising when one rewrites expressions in a form
mimicking SU(2) holonomies of the full theory, where one can choose an
irreducible representation for holonomies. Matrix elements in a
representation $j$ will then be of the form of exponentials used above
with exponents between $-j\mu_0c$ and $j\mu_0c$, the value in
(\ref{bracket}) corresponding to the fundamental representation
$j=\frac{1}{2}$. This procedure results in an effective density of the
form $d_l(p/p_*)=|p|^{-3/2}p_l(p/p_*)^{3/(2-2l)}$ with
\cite{Ambig,ICGC}
\begin{eqnarray} \label{pl}
 p_l(q) &=&
\frac{3}{2l}q^{1-l}\left((l+2)^{-1}
\left((q+1)^{l+2}-|q-1|^{l+2}\right)\right.\\
 &&\qquad- \left.(l+1)^{-1}q
\left((q+1)^{l+1}-{\rm sgn}(q-1)|q-1|^{l+1}\right)\right) \,. \nonumber
\end{eqnarray}

Similarly, the gravitational part of the Friedmann equation requires
reformulations because it contains terms linear and quadratic in $c$
while only exponentials of $ic$, i.e.\ almost periodic functions of
$c$, can be quantized. Thus, the classical expressions will be
replaced by a function which reduces to the classical one for small
$c$, i.e.\ small extrinsic curvature in a flat model, while giving
corrections at larger curvature. This usually also involves free
parameters as in the simplest case of choosing
\begin{equation} \label{holonomy}
 \mu_0^{-2}\sin^2\mu_0 c\sim c^2+O(c^4).
\end{equation}
In parallel with full constructions \cite{RS:Ham,QSDI}, this can be
understood as arising from the trace of a holonomy of $A^i_a$ around a
square loop whose edges are integral curves of symmetry generators
$X_I$. The square loop holonomy for edges along directions $I$ and $J$
is of the form
\begin{eqnarray*}
 h_Ih_Jh_I^{-1}h_J^{-1} &=& \cos^4(\case{1}{2}\mu_0c)+ 2
 (1+2\epsilon_{IJ}{}^K\tau_K) \sin^2(\case{1}{2}\mu_0c) 
\cos^2(\case{1}{2}\mu_0c)+
 (2\delta_{IJ}-1) \sin^4(\case{1}{2}\mu_0c)\\
 &&  + 4(\tau_I-\tau_J)
 (1+\delta_{IJ}) \sin^3(\case{1}{2}\mu_0c)\cos(\case{1}{2}\mu_0c)
\end{eqnarray*}
which, when appearing in a suitable trace, gives rise to a single term
$4\sin^2(\frac{1}{2}\mu_0c)
\cos^2(\frac{1}{2}\mu_0c)= \sin^2\mu_0c$ \cite{IsoCosmo}.
The parameter $\mu_0$ then determines the ratio of the coordinate
length of an edge of the loop to $V_0^{1/3}$, and it appears in the
holonomy through $\int_e\md tA_a^i\dot{e}^a\tau_i$ which has matrix
elements $\mu_0 V_0^{1/3} \tilde{c}/2$ for an isotropic connection. On
top of that, one can again choose non-fundamental representations for
holonomies with the effect of multiplying $\mu_0$ with a spin label
$j$ \cite{Gaul,AmbigConstr}.

In the Poisson brackets (\ref{bracket}) as well as in replacements
(\ref{holonomy}) for polynomials in $c$ we use holonomies such that it
appears most natural to use the same value for $\mu_0$ here. Its value
remains undetermined, but the order of magnitude has been estimated by
relating it to the lowest eigenvalue of the full area operator,
corresponding to the area of a square loop used in such a basic
dynamical move \cite{Bohr}. The argument is, however, incomplete
because the area operator is used to fix a parameter of the reduced
constraint although it does not appear in the full one as it is
formulated now. It seems against the general viewpoint of loop quantum
cosmology, formulated e.g.\ in \cite{LivRev}, to invoke the area
operator to quantize curvature components in a model while it does not
occur in the full constraint. Moreover, the coordinate area relevant
for the loop in holonomies is different from the geometrical area
quantized by the area operator.

These construction steps give rise to quantum corrections of different
types to classical equations, and also to several parameters to choose
which are ultimately to be related to the full theory. Open questions
in this context concern the relative magnitude of correction terms,
obtained through modifications such as those in (\ref{pl}) or
(\ref{holonomy}) in effective equations
\cite{Inflation,Closed,SemiClassEmerge,DiscCorr,Josh,Karpacz},
with respect to each other, which is important to know for the
construction of realistic scenarios, and natural ranges of the
parameters. While these open issues can be avoided to some degree by
sufficiently general arguments such as those for singularity removal
\cite{DegFull} or phenomenology \cite{Robust}, there are additional
problems some of which have become visible recently:
\begin{itemize}
 \item The parameter $j$ labels an SU(2) representation that needs to
 be chosen for the construction of inverse triad operators and is
 sometimes taken to be significantly larger than $1/2$ which would be
 the lowest allowed value corresponding to the fundamental
 representation. This permits one to remain well in a semiclassical
 regime while still having sizable quantum effects. It also justifies,
 when used only in inverse power corrections (\ref{inveigen}), to
 ignore other correction terms coming from holonomies in
 (\ref{holonomy}) because they are much smaller in such regimes.  It
 is thus meaningful for technical purposes to work with larger values
 of $j$ and analyze corresponding effects.  But it looks ``unnatural''
 that the representation label should be large, rather than just the
 value for the fundamental representation, or that different
 corrections should use different $j$. Moreover, there are arguments
 related to a type of quantum stability \cite{AmbigConstr} or the
 physical inner product \cite{AlexAmbig} which may indicate that the
 fundamental representation at least for holonomy corrections
 (\ref{holonomy}) is indeed preferred by internal consistency.%
\footnote{Those arguments
are, however, not as clear-cut as they are sometimes presented. As the
authors of \cite{AmbigConstr,AlexAmbig} mention, their
counter-arguments can easily be evaded by working not with a single
larger value of $j$ but by appropriate sums of operators involving
different $j$.}
\item In the construction of the dynamical equations one has to
replace ploynomials in $c$ by functions depending on
exponentials. This is justified when $c$ is small in a semiclassical
situation, while close to classical singularities there can be strong
differences between such functions which are indeed expected when
classical gravity breaks down. However, the isotropic connection
component $c$ can even become large in semiclassical regimes where one
would not expect strong quantum effects. This is the case whenever
there is a positive cosmological constant for which the
connection component behaves as $\tilde{c}\sim \sqrt{\Lambda}a$. Thus,
even if $\Lambda$ is small, $a$ will grow without bound and eventually
lead to a large $\tilde{c}$ (see Sec.~4.3 of \cite{IsoCosmo} and
\cite{ContFrac} for an early and a more recent discussion). Similar
effects, depending on initial values, may occur in anisotropic models
\cite{InstabAniso}.
\item In a flat model we have to choose a cell of coordinate size
$V_0$, or a compactification to a torus of the same size. This is just
auxiliary in flat models while its value is fixed in closed
ones. However, this value appears in equations through $d(p)$ and in
physical quantities such as the scale at which a bounce may occur
\cite{EffHam,GenericBounce,APS}. This is because the rescaled
variables $p$ and $c$, while they are not coordinate dependent, depend
on the value of $V_0$ entering in the rescaling.
\end{itemize}

These effects have been recognized and in some cases ameliorated by
amendments introduced in isotropic models. For instance, the issue in
the presence of a cosmological constant can be resolved if one chooses
$\mu_0$ not as a constant but as a $p$-dependent function $\bar{\mu}$
\cite{RSLoopDual,APS}. If, e.g., $\bar{\mu}(p)\sim 1/\sqrt{|p|}$ at
large scales, $c$ is always small in semiclassical regimes. Moreover,
such a choice can remove the $V_0$-dependence from the bounce scale
resulting from holonomy corrections \cite{RSLoopDual}. However, while
such a choice is not inconsistent with quantization procedures,
homogeneous models cannot justify it convincingly.
As with $\mu_0$, one is invoking the area operator although it does
not appear in current versions of the full constraint.  Moreover, the
resulting operator for curvature components becomes triad-dependent
which is not realized in the full theory either. We will demonstrate
in what follows that the intuition behind those modifications is
nonetheless borne out by lattice constructions.

In fact, isotropic loop quantum cosmology was originally developed for
models which exhibit typical effects close to classical singularities
in order to see how quantum corrections can lead to better classical
behavior.  In the meantime, the resulting singularity removal
mechanism has been extended non-trivially to inhomogeneous models
\cite{SphSymmSing}.  This was successful because for this aspect a
single spatial patch is already quite typical and thus homogeneous
models are reliable (while isotropy itself would be too special; see
e.g.\ \cite{DegFull}). These models were not intended to be taken too
seriously for quantitative aspects at larger scales. Although the
universe is homogeneous on such scales to a high degree, the classical
continuum picture requires space to be made of many ``atoms'' in a
discrete world. Thus, at the quantum level one must not ignore
inhomogeneities especially when the universe becomes large. Some of
the above difficulties arise precisely from taking isotropic models
literally at large scales. One can clearly see this for the
cosmological constant, which implies a large integrated trace of
extrinsic curvature just because space becomes large.  This is the
case even if the local curvature scale remains small in a
semiclassical regime. While homogeneous models do not allow much
choice in basic variables, which will include some form of the total
extrinsic curvature, inhomogeneous models are built on more local
objects.  A better situation can then be expected if inhomogeneous
models are used where basic variables remain small in semiclassical
regimes.

\section{Inhomogeneous effects}

The full theory in current form has several different complications
not all of which are related to inhomogeneities as they arise as modes
on a background.

\subsection{Aspects of loop quantum gravity}

Configuration variables are holonomies associated with arbitary curves
in space, whose values allow one to distinguish between all
connections relevant for general relativity. Since the connection
takes values in su(2), holonomies are elements of the group
SU(2). Their matrix elements are multiplication operators in the basic
representation of loop quantum gravity built on states which are
gauge invariant products of holonomies associated with edges of
arbitrary spatial graphs \cite{ALMMT}. While these graphs visualize
the discrete spatial structure of quantum geometry, they can be
arbitrarily fine and no explicit short-scale cut-off is present in the
theory. Graphs can also be of any topology (knotted or linked) and can
have vertices of high valence. While holonomies implement the
connection as basic operators, spatial geometry is given by triads, the
conjugate momenta. They are quantized to flux operators, associated
with 2-surfaces in space, which take non-zero contributions only from
intersection points between the surface and edges of the graph acted
on by the flux operator. Their spectra are discrete, further
implementing the spatial discreteness of quantum geometry.

These basic operators appear in more complicated ones relevant for
dynamics, such as matter Hamiltonians for which inverse triad
operators are necessary and the gravitational Hamiltonian
constraint. While such operators can be constructed in well-defined
manners \cite{QSDI,QSDV}, their actions are highly involved on
arbitrary states. Moreover, in particular the gravitational constraint
usually changes the graph underlying a state it acts on because
holonomies around closed loops are used to quantize curvature
components. Unless such loops are lying entirely on the original graph
of a state, the action creates new edges and new vertices leading to a
finer graph. It is a success, however, that such operators can be
defined in a well-defined manner at all, considering that their
analogs in quantum field theory on a curved background space-time
would suffer from several infinities. This is one of the places where
background independence and the quantum representation it leads to are
crucial and imply characteristic properties of resulting
theories. As noted before, there is no explicit short-scale cut-off in
the theory. Finiteness rather results from the fact that Hamiltonian
operators act on graph states, and each such graph implies a non-local
representation of the classical fields.

In addition to the Hamiltonian constraint, there is the diffeomorphism
constraint which is implemented as finite diffeomorphisms moving
graphs in space. Solving it implies that invariant results are
independent of the embedding of graphs in space. Most or all of this
freedom needs to be fixed when calculations are to be done using a
background geometry for a particular physical regime. In the full
setting, this can be achieved by picking suitable semi-classical
states peaked at a geometry corresponding to the desired background
(see, e.g., \cite{CohState}). In this way, the full background
independent framework is used but special states are selected to pick
a regime, making use of an additional (background) structure. In
addition to the classical background, such states depend on typical
quantum aspects such as the spread of a state or parameters describing
the typical fineness of graphs used and thus encoding discrete spatial
aspects.

Unfortunately, there are conceptual and technical difficulties in even
defining suitable semiclassical states, and working with them at this
level is highly involved. One of the difficulties is, for instance,
that holonomies are SU(2)-valued which requires the use of lenghty
re-coupling identities in doing explicit calculations. On the other
hand, results are rarely sensitive to all aspects and can often be
reproduced with a high level of accuracy in simpler constructions
(compare, e.g., \cite{CorrectScalar} with \cite{QFTonCSTII}). We will
now devise a model which allows one to formulate a suitably explicit
framework for perturbative inhomogeneities and other inhomogeneous
regimes such as the BKL picture, maintaining the characteristic properties
of a background independent quantization.

\subsection{Lattice models}

To capture significant inhomogeneous ingredients we
introduce regular lattice states with spacing $\ell_0$ measured in
background coordinates assumed to be flat. 
The motivation is to analyze implications of typical aspects of the
full theory such as the fact that states are based on graphs and
properties of Hamiltonian operators. This will allow one to read off
characteristic modifications to classical expressions which can then
be transferred to effective equations.
Lattice links are parts of integral curves of a basis of symmetry
generators of the background, which also provides an orientation of
all edges. Although it is not necessary for inhomogeneous states, we
assume this lattice to be in a cell of finite size $V_0$ for
comparison with homogeneous models. The number of lattice blocks and
vertices is then $N=V_0/\ell_0^3$. Lattices of this type are special
cases of fundamental states, but we also make use of background
structures to be introduced in the construction. This renders the
usual floating lattice of a background independent fundamental
description into a rigid one on a background, suitable, e.g., for
cosmological considerations. By explicitly introducing a background in
this manner we bypass more involved reformulations which would
define a background through relational objects, akin to relational
solutions of the problem of time.

Usually, spin network states are built on graphs, labeled by SU(2)
representations on edges and contraction matrices in vertices in order
to multiply all holonomies, evaluated in the labeling representation,
to a gauge invariant function of the connection. This changes when the
construction is used on a background, for instance a homogeneous one
where integral curves of Killing vector fields define the lattice
links. For a perturbative treatment of inhomogeneities classical
variables of the form $E^a_i=\tilde{p}^I(x)\delta^{(i)}_I\delta^a_i$
and $A_a^i=\tilde{k}_I(x) \delta^I_{(i)}\delta_a^i+
\psi_I(x)\epsilon^{Ii}_a$ are suitable where the densitized triad and
the first part of the connection are diagonal.  A diagonal densitized
triad is obtained by a gauge choice (such as scalar modes in
longitudinal gauge) making use of the background around which one
perturbs. The connection cannot be completely diagonal, however,
because this would violate the Gauss constraint in an inhomogeneous
situation (see also \cite{SphSymm,SphSymmVol}). The non-diagonal part
containing $\psi_I(x)$ comes from the spin connection which is
determined completely in terms of $p^I(x)$, and possibly spatial
derivatives of the shift vector if it is not zero in the chosen
gauge. The independent degree of freedom $\tilde{k}_I(x)$ in the
connection thus appears in diagonal form (in gauges where the shift
vector vanishes this corresponds to extrinsic curvature; otherwise
there is an additional contribution to the non-diagonal connection
part).  In fact, we have conjugate fields $\tilde{k}_I(x)$ and
$\tilde{p}^I(x)$, with
\begin{equation}
 \{\tilde{k}_I(x),\tilde{p}^J(y)\} = 8\pi\gamma
 G\delta_I^J\delta(x,y)\,.
\end{equation}
Taking only the diagonal part of $A_a^i$, holonomies along links are
of the form $\exp(\ell_0\tilde{k}_I\tau_I)$ in direction $I$. The
diagonal form of basic variables implies that not all the freedom of
an SU(2) theory has to be dealt with, effectively Abelianizing the
framework as in homogeneous models
\cite{cosmoI,cosmoII,DegFull}. Nevertheless, we can consider higher
representations of SU(2) such that, in the $j$-representation, a
holonomy has matrix elements $\exp(ij\ell_0\tilde{k}_I)$ and lower
powers. Independent functions on the space of lattice connections are
thus labeled by integers $\mu_{I,\vertex}$ where $\vertex$ denotes the
vertex position and $I=1,2,3$ the direction of the link starting from
the vertex using the orientation of symmetry generators as tangent
vectors to links. Such functions are then of the form
$\exp(i\mu_{I,\vertex} \ell_0 \tilde{k}_I/2)$. In homogeneous models,
one uses real labels $\mu$ for representations of the Bohr
compactification of the real line, which can be seen as arising from a
degeneracy between representation labels and edge length
\cite{Bohr}. The functions $\exp(i\mu c/2)$ are then used in isotropic
models, which separate the space of isotropic connections. Using
integer labels on a fixed lattice states does not allow us to separate
all classical connections, but this is not required because choosing a
lattice means that only functions of a certain scale are being probed.

At this point, we have assumed diagonal metrics also at the
inhomogeneous level which effectively Abelianizes the theory: instead
of SU(2) calculations we simply work with U(1), i.e.\ we can replace
complicated SU(2) recoupling relations by multiplications of
phases.\footnote{As stated above, this is sufficient for a
perturbative treatment of inhomogeneities.}  This is the reason why we
do not need to introduce vertex labels because Abelian holonomies are
uniquely multiplied to gauge-invariant functions. Although the
diagonalization is a truncation of the classical theory, it allows
full access to perturbative inhomogeneous degrees of freedom realized
classically. This is also the key reason for simplifications that
makes it possible to do explicit calculations.

\begin{figure}
\centerline{\includegraphics[width=7cm,keepaspectratio]{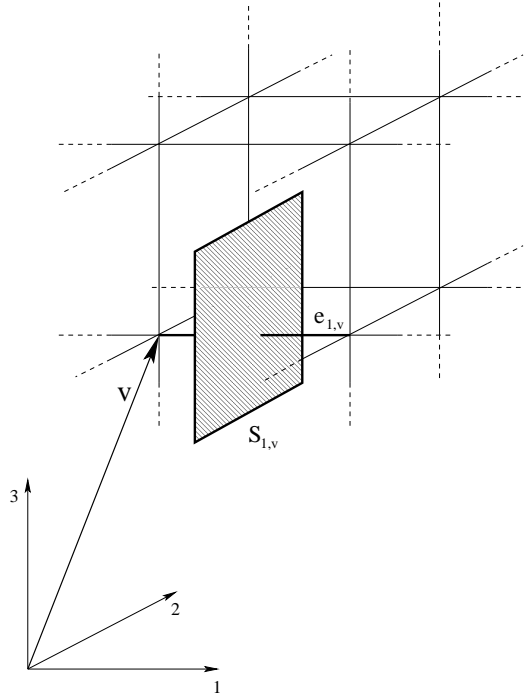}}
\caption{Lattice with elementary edge and surface. \label{flux}}
\end{figure}

Since this corresponds to a field theory for classical variables
$\tilde{p}^{I}(x)$ and $\tilde{k}_I(x)$, we have many more basic
operators which are nevertheless constructed very similarly to those
of isotropic models. Holonomies along lattice links $e_{I,\vertex}$ are
of the form
\[
 h_{I,\vertex}=\exp (i\smallint_{e_{I,\vertex}}\md t 
\tilde{k}_I/2)\approx
\exp(i\ell_0\tilde{k}_{I}(\vertex)/2)=\exp(ik_I(\vertex)/2)
\]
absorbing $\ell_0$ in $k_I:=\ell_0\tilde{k}_I$,
and fluxes through lattice sites $S_{I,\vertex}$ perpendicular to an
edge $e_{I,\vertex}$ as in Fig.~\ref{flux} are given by $F_{I,\vertex}=
\int_{S_{I,\vertex}} \tilde{p}^I(y)\md^2y=
\ell_0^2\tilde{p}^I(\vertex)=: p^I(\vertex)$. Compared to isotropic
models, $V_0^{1/3}$ is thus replaced by $\ell_0$ and the lattice
spacing takes over the role of the cell size. Now all expressions are
dependent on $\ell_0$, but this has meaning as the lattice size,
although it is just the size measured in background
coordinates. Nonetheless, $\ell_0$ is not just an auxiliary quantity
because it is thought of as arising from a fundamental state after
re-introducing a background rather than being introduced to discretize
a theory by hand.

Poisson relations between holonomies and fluxes introduced here
follow, e.g., from a mode decomposition
\begin{eqnarray*}
 \tilde{k}_I(x) &=&\sum_{\kappa} \tilde{k}_I(\kappa)e^{i\kappa\cdot x}\\
 \tilde{p}^J(x) &=&\sum_{\kappa} \tilde{p}^J(\kappa)e^{i\kappa\cdot x}
\end{eqnarray*}
of the fields $\tilde{k}_I(x)$ and $\tilde{p}^J(x)$ on a symmetric
background which we still assume to be flat for simplicity. With our
box of size $V_0$, all components of the wave numbers $\kappa$ summed
over are $\kappa_I=2\pi V_0^{-1/3}n$ with positive integers $n$. From
\[
 \int \dot{\tilde{k}}_I(x)\tilde{p}^I(x)\md^3x = \sum_{\kappa,\kappa'}
\dot{\tilde{k}}_I(\kappa)\tilde{p}^I(\kappa') 
\int e^{i(\kappa+\kappa')\cdot x}\md^3x =
V_0 \sum_{\kappa} \dot{\tilde{k}}_I(\kappa)\tilde{p}^I(-\kappa)
\]
we obtain Poisson brackets
\[
 \{\tilde{k}_I(\kappa),\tilde{p}^J(\kappa')\} =8\pi\gamma
GV_0^{-1}\delta_I^J\delta_{\kappa,-\kappa'}
\]
between the modes. Link integrals of the connection and fluxes then
become
\begin{eqnarray*}
 {\cal I}_I(\vertex) &:=& \int_{e_{I,\vertex}}\md t \tilde{k}_I(e(t)) =
\int_{e_{I,\vertex}}\md t\sum_{\kappa}\tilde{k}_I(\kappa)e^{i\kappa\cdot
e(t)}= \sum_{\kappa}
\tilde{k}_I(\kappa)e^{i\kappa\cdot\vertex} \int_0^{\ell_0}e^{i\kappa_It}\md t\\
 &=& 2\sum_{\kappa}\tilde{k}_I(\kappa)e^{i\kappa\cdot\vertex}
e^{i\kappa_I\ell_0/2}\sin(\kappa_I\ell_0/2)/\kappa_I
\end{eqnarray*}
and
\begin{eqnarray*}
 F_{J,\vertexalt} &=& \int_{S_{J,\vertexalt}}\md^2y
\sum_{\kappa}\tilde{p}^J(\kappa)e^{i\kappa\cdot y}= 
\sum_{\kappa}\tilde{p}^J(\kappa)
e^{i\kappa\cdot\vertexalt}  e^{i\kappa_J\ell_0/2}
\int_{-\ell_0/2}^{\ell_0/2} e^{i\kappa_Kt}\md t
\int_{-\ell_0/2}^{\ell_0/2} e^{i\kappa_Lt}\md t\\
&=&4\sum_{\kappa}\tilde{p}^J(\kappa)e^{i\kappa\cdot\vertexalt} 
e^{i\kappa_J\ell_0/2}
\sin(\kappa_K\ell_0/2)\sin(\kappa_L\ell_0/2)/\kappa_K\kappa_L
\end{eqnarray*}
where the values of indices $K$ and $L$ are defined such that
$\epsilon_{JKL}=1$.  Thus,
\begin{eqnarray}
 \{{\cal I}_{I,\vertex},F_{J,\vertexalt}\} &=& 64\pi\gamma GV_0^{-1}\delta_I^J
\sum_{\kappa} e^{i\kappa\cdot(\vertex-\vertexalt)}
\sin(\kappa_I\ell_0/2)\sin(\kappa_K\ell_0/2)\sin(\kappa_L\ell_0/2)/
(\kappa_I\kappa_K\kappa_L)\nonumber\\
&=&8\pi\gamma G\delta_I^J\chi_{\ell_0}(\vertex-\vertexalt)= 8\pi\gamma
G\delta_I^J\delta_{\vertex,\vertexalt}\,.
\end{eqnarray}
Using Fourier series leads to the appearance of characteristic
functions $\chi_{\ell_0}(x)$ centered at $x$ of width
$\ell_0$. Restricted to vertices on a lattice of spacing $\ell_0$,
this is identical to $\delta_{x,0}$.

When quantized, holonomies will again become multiplication operators
\begin{equation}
 \hat{h}_{I,\vertex}|\ldots,\mu_{I,\vertex},\ldots\rangle =
|\ldots,\mu_{I,\vertex}+1,\ldots\rangle
\end{equation}
and change the link labels when acting on a state. Note that we do not
use a parameter like $\mu_0$ since edges of the link size are
distinguished and multiplying with holonomies for edges not integer
multiples of lattice links does not preserve the lattice
structure. The parameter $\mu_0$ is thus an integer, simply
corresponding to the choice of representation of holonomies. For basic
holonomies, $\mu_0=1$.

Flux operators
\begin{equation}
 \hat{F}_{I,\vertex} |\ldots,\mu_{I,\vertex},\ldots\rangle= 
 \frac{1}{2}\gamma\lP^2\mu_{I,\vertex}
|\ldots,\mu_{I,\vertex},\ldots\rangle
\end{equation}
corresponding to a lattice site orthogonal to an edge in direction $I$
have eigenvalues proportional to $\mu_{I,\vertex}$ or, when we take a
surface intersected by the vertex $\vertex$, proportional to the
average of labels of neighboring edges,
\begin{equation}
 \hat{{\cal F}}_{I,\vertex} |\ldots,\mu_{I,\vertex},\ldots\rangle= 
 \frac{1}{4}\gamma\lP^2(\mu_{I,\vertex+e_{I,\vertex}}+\mu_{I,\vertex})
|\ldots,\mu_{I,\vertex},\ldots\rangle\,.
\end{equation}
In this notation, $\vertex+e_{I,\vertex}$ denotes the vertex next to
$\vertex$ along the edge $e_{I,\vertex}$, i.e.\ the endpoint of
$e_{I,\vertex}$ other than $\vertex$.
These labels then determine the volume eigenvalues. From
\[
 V=\int\md^3x\sqrt{|\tilde{p}^1\tilde{p}^2\tilde{p}^3|}\approx
\sum_{\vertex} \ell_0^3
\sqrt{|\tilde{p}^1(\vertex)\tilde{p}^2(\vertex)\tilde{p}^3(\vertex)|}=
\sum_{\vertex} \sqrt{|p^1(\vertex)p^2(\vertex)p^3(\vertex)|}
\]
the volume operator is defined as $\hat{V}=\sum_{\vertex} \prod_{I=1}^3
\sqrt{|\hat{\cal F}_{I,\vertex}|}$ with eigenvalues
\begin{equation}
 V(\{\mu_{I,\vertex}\})=
\left(\frac{\gamma\lP^2}{4}\right)^{3/2} 
 \sum_{\vertex} \prod_{I=1}^3\sqrt{
|\mu_{I,\vertex}+\mu_{I,\vertex+e_{I,\vertex}}|}\,.
\end{equation}

Also as in isotropic models, we can construct composite operators such
as those for inverse triad components. The only difference is that we
construct local versions of such operators localized at vertices, for
which we use neighboring link holonomies. Eigenvalues of such
commutators will then look similar to those in the isotropic case,
except that we have a sum over vertex contributions where single link
labels change by $\pm1$ in the form $V_{\vertex}(\mu_{I,\vertex}+1)-
V_{\vertex}(\mu_{I,\vertex}-1)$, rather than one global contribution as
in (\ref{inveigen}). Also, as noted before these functions do not
contain $\mu_0$ since the link size is fixed. By the same calculations
used to derive $d(p)$ in the isotropic case, such functions have peaks
at values $\mu_*\approx1$, or $\mu_*\approx j$ for an arbitrary
irreducible representation, corresponding to
$p_*\approx\frac{1}{6}\gamma j\lP^2$. This is similar to the isotropic
case, but now $p_*$ distinguishes different regimes according to
$\ell_0^2\tilde{p}>p_*$ or $\ell_0^2\tilde{p}<p_*$ rather than
inequalities for the isotropic $p=V_0^{1/3}\tilde{p}$. So also here,
$V_0$ has been replaced by the smaller $\ell_0$.

\section{Basic observations}

Having reviewed isotropic basic variables and introduced analogous
ones for inhomogeneous lattice states we can draw conclusions
regarding their relation.  Most of these effects are qualitatively
true for any inhomogeneities, but are made explicit in lattice models
introduced here.

\subsection{Rescaling freedom}

As noted before, coordinate background structures to define
homogeneity or the choice of lattices in inhomogeneous states introduce
parameters such as the cell volume $V_0$ or the lattice spacing
$\ell_0$. The parameters then occur in derived expressions, and such
expressions change when other choices for the parameters are
made. Since the parameters are related to coordinates or the embedding
of structures in space, it is not always guaranteed that such a
rescaling freedom makes physical sense. Indeed, for $V_0$ this is not
the case which is one of the difficulties in purely homogeneous
models. Ratios such as $\mu/\mu_*=p/p_*=V_0^{2/3}\tilde{p}/p_*$ which
demarcate classical from quantum behavior, where $\mu_*$ is a
characteristic scale appearing in the construction of operators and
$\mu$ a state label, depend on $V_0$ through $p$. (Factors of $V_0$ do
not cancel since $p_*$ is defined independently of $V_0$.)

The situation is, however, different for $\ell_0$ which replaces $V_0$
in elementary expressions when lattice states are used. Although $V_0$
is still present, it always appears in combination with $\ell_0$
through $N$. Then, ratios as before take the form
$\mu/\mu_*=p/p_*=\ell_0^2\tilde{p}/p_*$ which is
$\ell_0$-dependent. But unlike $V_0$, changing $\ell_0$ has physical
meaning because we then change the scale on which we probe space by
the lattice. If we choose a bigger lattice spacing, i.e.\ enlarge
$\ell_0$, $\tilde{p}$ will have to drop to even smaller values for
quantum corrections to become noticeable. Since both $\ell_0$ and
$\tilde{p}$ are coordinate dependent, this statement about the
relation between changing $\ell_0$ and corresponding changes in
$\tilde{p}$ is invariant under rescaling of coordinates. This happens
in such a way that the change of $\ell_0$ implies physical
properties that are expected. Isotropic models show similar
technical features, but due to additional backgrounds involved they
are not as physically convincing.

These observations also indicate that the triad scale at which quantum
effects become important in inverse powers does not only depend on a
representation label chosen for the quantization of inverse triad
operators but also on the lattice size. In fact, choosing finer
lattices has the same effect as choosing higher representations on a
lattice of unchanged size. The importance of such effects is
determined by ratios $p/p_*$ where $p$ is a flux value and $p_*$ is
proportional to $j$. Technically, this ratio can be made small by
choosing larger $j$, or by making the relevant fluxes take on smaller
values in the same situation. In homogeneous settings one only has the
first option, while inhomogeneous ones easily allow the second one by
choosing smaller surfaces for integrating fluxes.

In fact, if the geometry is nearly isotropic, we can use the average
value $p=V_0^{2/3}\tilde{p}$ to make contact with isotorpic
variables. Now, as discussed at the end of the preceding section,
quantum corrections start to become noticeable when
$\ell_0^2\tilde{p}\approx p_*$, or
$p=V_0^{2/3}\tilde{p}=N^{2/3}\ell_0^2\tilde{p}\approx N^{2/3} p_*$
which is enhanced by a factor depending on the number of
vertices. Note that we did not use higher representations for
holonomies in commutators, by which one can achieve a similar effect
if $j\sim N^{2/3}$. The enhancement comes just from the fact that for
individual links it is the local value of $\tilde{p}$ rather than the
total one which is relevant.

\subsection{Higher spin representations}

This observation brings us to the next point, which is the naturalness
or unnaturalness, or even consistency, of using higher representations
of holonomies to construct operators. As we have seen, in
inhomogeneous systems one can achieve the same effect by using finer
lattices which is certainly a legitimate way to change
parameters. While this is not available in homogeneous models, using
higher spins there may simply be seen as a way to mimic inhomogeneous
behavior in such a setting. There is no direct relation between all
features of homogeneous models to the full theory because of
degeneracies: changing very different ingredients of an inhomogeneous
situation can result in the same change in a homogeneous model. The
higher spin compared to the lattice size is one such example, where in
earlier papers only the direct relation identifying a higher
representation in a homogeneous model with a higher representation in
the full theory has been made. If this is the only way to relate
higher spins to properties of the full theory, it certainly makes
higher spins in homogeneous models look unnatural. The relation to the
lattice spacing, overlooked so far, makes using higher representations
much more natural in homogeneous models, while inhomogeneous ones can
still be formulated by restricting oneself to the fundamental
representation in composite operators.

Along similar lines one can justify using different spins for
gravitational and matter parts in a Hamiltonian constraint as
effective means to include inhomogeneous features.  Even if we use
higher spins in homogeneous models, one may still question why one
should use different ones in holonomies in a matter Hamiltonian
compared to the gravitational part of the constraint. Such different
spins are often helpful to bring out matter corrections as the
dominant ones since they are most easy to derive and to work with. The
relation to lattice spacing suggests a reason for higher spins in
matter terms compared to gravitational ones: it simply means that for
matter terms we use longer range ``interactions'' based on holonomies
which extend through several vertices rather than just one basic
lattice link. This is an option one clearly has in any lattice model,
for which loop quantum gravity is one example. If one accepts such
longer range interactions, one is led directly to the behavior also
resulting from higher spins in a homogeneous model.

\subsection{Cosmological constant}

In addition to ratios $p/p_*$ relevant for inverse triad operators,
connection components entering holonomies are the second variable
which determines where quantum corrections become noticeable. In
isotropic models, $\mu_0 c$ has to be small compared to one for the
classical constraint as a polynomial in $c$ to be a good approximation
of functions of exponentials
$\exp(i\mu_0c/2)=\exp(i\mu_0V_0^{1/3}\tilde{c}/2)$ used for the
quantization. As noted before, however, in the presence of a
cosmological constant $c$ can become arbitrarily large even in
classical regimes (while $\mu_0$ has been argued to be of order one by
comparing with the lowest area eigenvalue \cite{Bohr}).

In inhomogeneous lattice models, holonomies appearing in composite
operators are associated with links and of the form
$\exp(i\ell_0\tilde{c}/2)=\exp(iN^{-1/3}c/2)$ where we introduced the
isotropic $c=V_0^{1/3}\tilde{c}$ for comparison. What has to be small
is thus not $\mu_0c$ but $N^{-1/3}c$ which can be achieved by having a
large number of vertices even if $c$ does not become small. A large
number of vertices is in fact what one expects if a universe grows to
larger size because the fundamental Hamiltonian of loop quantum
gravity most likely creates new vertices in its action
\cite{RS:Ham,QSDI}. The creation of new vertices is not easily
incorporated in a regular lattice model, although it is possible, but
one can view regular lattices as a coarse-grained, effective
description of the fundamental one.  For this, no explicit mechanism
is required: for effective equations we can compute expansion
coefficients locally on a given lattice with $N$ vertices. In those
coefficients, metric values appear depending on $N$. Parametrically,
$N$ can then be assumed to be a function of the total volume or scale
factor whose precise behavior is to be determined by a more detailed
formulation. This is analogous to a mean field approximation where
``fast'' modes are treated only effectively by correction terms. Here,
vertices are created at each step of the action of the constraint
operator implying Planck-size changes in the local geometry. This
degree of freedom is thus much faster than semiclassical effects we
are mainly interested in here, i.e.\ it happens on much smaller time
scales (measured in terms of the global geometry).

Then, as the total volume increases one should also adapt the lattice
size and increase $N$. This is in particular the case when one has a
positive cosmological constant because the universe grows without
bound. As the scale factor $a$ increases and
$\tilde{c}\propto\sqrt{\Lambda}a$ increases, also $N$ has to
increase. For a well-behaved semiclassical state this has to happen in
such a way that $c/N^{1/3}$, the argument of holonomies, remains small
which is easily achieved in an inhomogeneous model. The validity of a
homogeneous model, on the other hand, will break down at a certain
volume, which is one example where homogeneous approximations are good
on small scales but not at large ones.\footnote{On very small scales
close to classical singularities, of course, inhomogeneities grow and
homogeneous models can only give qualitative insights. Classically,
one would expect better and better behavior of a homogeneous model as
an approximation to inhomogeneous behavior the larger the universe
grows. This is not realized in quantized models with a discrete
spatial structure because more and more elementary blocks are needed
to describe a large space. Very close to classical singularities, on
the other hand, most indications so far show that anisotropies rather
than inhomogeneities are the crucial degrees of freedom.}

We can draw one further conclusion related to this observation. In
link holonomies we have the quantity $N^{-1/3}c$ compared to the
isotropic $\mu_0c$ with a constant $\mu_0$. In inhomogeneous
situations, $\mu_0$ does not appear at all, and there is thus no
indication for the makeshift argument relating $\mu_0$ to the area
spectrum used in the absence of a relation to inhomogeneity.
This agrees with the fact that the full constraint operator does not
contain the area operator. Curvature components of the constraint are
rather quantized directly through holonomies, which is also realized
here with natural loops provided by the lattice.
Moreover, if $N^{1/3}$ is not treated as a constant, we can resolve
problems at large scales. This can be modeled in a purely isotropic
situation by letting $\mu_0$ be scale-dependent in a form
$\bar{\mu}(p)\sim |p|^{-1/2}$ at large scales, corresponding to the
number of vertices growing linearly with volume, $N\propto
|p|^{3/2}$. This is exactly the behavior which has been proposed
within isotropic models to counter large scale and other problems
\cite{RSLoopDual,APS}. While one might question a non-constant $\mu_0$
in isotropic models on the grounds that holonomies would have to
depend on triad variables, the relation to the number of vertices in
inhomogeneous states clearly provides convincing justification for it.
In general, however, the behavior may differ from $N\propto |p|^{3/2}$
which is to be checked in detailed models implementing the
subdivision. There are also conceptual differences between the
arguments for $\bar{\mu}(p)$ in isotropic models and those presented
here: while in isotropic models the parameter arises at the operator
level and leads to holonomies in the constraint depending on triad
components, lattice models implement a volume dependence through
states. This is closer to the full theory where it is the
state-dependent regularization of the Hamiltonian constraint that
determines how new edges and vertices are created. Nevertheless, it is
promising and encouraging for other constructions in models that
independent considerations in homogeneous and inhomogeneous settings
lead to the same qualitative conclusions.

\subsection{Relative size of quantum corrections}

Finally, we discuss the relative size of different quantum corrections
arising from quantum Hamiltonians. The main corrections are from
inverse triad operators, where the size of $p$ is relevant, and
holonomies where $c$ is relevant. While $p$ has to be sufficiently
{\em small} for inverse triad corrections to become large, $c$ has to
be {\em large} for large holonomy corrections. For evaluations of
effective equations
\cite{Inflation,DiscCorr,Josh,EffAc,Karpacz,Perturb} it is then
helpful to know if there is any correction which dominates, or if all
of them have to be taken together which would make the analysis more
complicated. In any model, an answer to this question depends on which
regime one is looking at, and the situation in homogeneous models is
undecided. Irrespective of what the answer in a particular homogeneous
scenario is, however, the situation is different in inhomogeneous
models. As before, inhomogeneity implies that for local operators
appearing in composite ones connections are integrated only over small
links and triads only over lattice sites rather than all of
space. Thus, the connection values relevant for holonomies and the
triad values for fluxes are both significantly reduced compared to the
values that appear in homogenous operators. As a consequence, inverse
triad corrections will become {\em more prominent} while holonomy
corrections will be {\em less so}. Main effects are then to be
expected from inverse triad corrections of quantum geometry, as they
have been studied several times in homogeneous models (see, e.g.,
\cite{Inflation,BounceClosed,BounceQualitative,Oscill,LoopFluid,NonChaos,Cyclic,Collapse}).

\section{Further applications}

Besides these technical observations we can also draw more general
conclusions.

\subsection{Symmetry reduction: from inhomogeneity to homogeneous
models}
\label{SymmRed}

Symmetric states are obtained by restricting full states to a subset
of invariant connections relevant for a symmetric model
\cite{SymmRed}. Such states are necessarily distributional in the
less symmetric situation, and general operators will not leave the
space of such states invariant. Nevertheless, one can derive all basic
operators given by holonomies and fluxes of the symmetric model from
corresponding ones in the full theory. This is sufficient to derive
the quantum representation of models from the unique one in the full
theory \cite{LOST,WeylRep}. Since many properties of loop
quantizations, including those discussed here, follow already from
basic aspects the derivation of the basic representation is a crucial
step. Explicit examples for such constructions can be found for
reducing an anisotropic model to an isotropic one in \cite{AnisoPert}
and for obtaining the spherically symmetric representation from the
full one in \cite{SphSymm}. A more general discussion is given in
\cite{LivRev}.

We can use the constructions presented here to shed more light on the
role of inhomogeneities in these reductions. A general lattice state
is of the form
\[
 \psi_{\{\mu_{I,\vertex}\}}[h_{I,\vertex}]= \prod_{I,\vertex}
h_{I,\vertex}^{\mu_{I,\vertex}}=
\langle k_J(x)|\ldots,\mu_{I,\vertex},\ldots\rangle
\]
with $k_I(x)=\ell_0\tilde{k}_I(x)$ and
$h_{I,\vertex}\approx\exp(i\mu_{I,\vertex}k_I(\vertex)/2)$. This
corresponds to an isotropic connection $\tilde{c}$ if
$\tilde{k}_I(x)=\tilde{c}$, or $k_I(x)=\ell_0V_0^{-1/3}c= N^{-1/3}c$,
for all $I$ and $x$. The restriction of the state then becomes
the isotropic state
\begin{equation}\label{restricted}
 \psi_{\mu}(c)=\exp(i\mu c/2)=:\langle c|\mu\rangle \quad\mbox{with}\quad
\mu=N^{-1/3}\sum_{I,\vertex}\mu_{I,\vertex}\,.
\end{equation}
Following the general procedure, there is a map $\sigma$ taking an
isotropic state of the reduced model to a distributional state
$(\mu|=\sigma(|\mu\rangle)$ in the inhomogeneous setting such that
\begin{equation}
 (\mu|\ldots,\mu_{I,\vertex},\ldots\rangle=
\langle\mu|\ldots,\mu_{I,\vertex},\ldots
\rangle|_{\tilde{k}_{I}(x)=\tilde{c}} \quad\mbox{for
all}\quad |\ldots,\mu_{I,\vertex},\ldots\rangle
\end{equation}
where the inner product on the right hand side is taken in the
isotropic Hilbert space, using the restricted state
$|\ldots,\mu_{I,\vertex},\ldots\rangle|_{\tilde{k}_{I}(x)=\tilde{c}}$ as an
isotropic state (\ref{restricted}).

Link holonomies as multiplication operators simply reduce to
multiplication operators on isotropic states. Fluxes for lattice
sites, however, do not map isotropic states to other isotropic
ones. This can easily be seen using $(\mu|\hat{F}_{J,\vertexalt}
|\ldots,\mu_{I,\vertex},\ldots\rangle= \frac{1}{2}\gamma\ell_{\rm
P}^2\mu_{J,\vertexalt} |\ldots,\mu_{I,\vertex},\ldots\rangle$ on states 
\[
 |\psi_{I,\vertex}\rangle:=|0,\ldots,0,1,0,\ldots,0\rangle
\]
which have non-zero labels only on one lattice link $e_{I,\vertex}$. We
then have
\[
 (\mu|\hat{F}_{I,\vertex}|\psi_{I,\vertex}\rangle= \frac{1}{2}\gamma\ell_{\rm
P}^2\mu_{I,\vertex} \delta_{\mu,1} \quad\mbox{and}\quad
(\mu|\hat{F}_{I,\vertex}|\psi_{I,\vertex+e_{I,\vertex}}\rangle=0
\]
since the flux surface and the non-trivial link do not intersect in
the second case. However, $(\nu|\psi_{I,\vertex}\rangle=
(\nu|\psi_{I,\vertex+e_I}\rangle$ for any isotropic state
$(\nu|$. Thus, $(\mu|\hat{F}_{I,\vertex}$ cannot be a superposition of
isotropic distributional states, and flux operators associated with a
single link do not map the space of isotropic states to itself. (The
above formulas show that $(1|$ cannot be contained in a decomposition
of $(\mu|\hat{F}_{I,\vertex}$ in basis states, but we can repeat the
arguments with arbitrary values for the non-zero label in
$|\psi_{I,\vertex}\rangle$.)

Instead, one can use extended fluxes which are closer to homogeneous
expressions. First, we extend a lattice site flux $\hat{F}_{I,\vertex}$ to
span through a whole plane in the lattice, leading to
$\sum_{\vertexalt:\vertexalt_I=\vertex_I}\hat{F}_{I,\vertexalt}$.

This corresponds to a homogeneous flux in the box of size $V_0$ but is
still not translationally invariant because the plane
$\{\vertexalt:\vertexalt_I=\vertex_I\}$ is distinguished. We can make it
homogeneous on the lattice by averaging along the direction $I$
transversal to the plane. This leads to a sum over all lattice
vertices in
\[
 \hat{p}^I:=N^{-1/3}\sum_{\vertex} \hat{F}_{I,\vertex}
\]
including a factor $N^{-1/3}$ from averaging in one direction.
Finally, we take the directional average
\[
 \hat{p}=\frac{1}{3}\sum_I\hat{p}^I=\frac{1}{3N^{1/3}}\sum_{I,\vertex}
\hat{F}_{I,\vertex}
\]
to define the isotropic flux operator.

Now, to find the action of this operator on a distributional state
$(\nu|$ we compute
\[
 (\nu|\hat{p}|\ldots,\mu_{I,\vertex},\ldots\rangle=
\frac{\gamma\ell_{\rm P}^2}{6N^{1/3}}\sum_{J,\vertexalt}
\mu_{J,\vertexalt} (\nu|\ldots,\mu_{I,\vertex},\ldots\rangle
=\frac{1}{6}\gamma\ell_{\rm P}^2\mu \delta_{\nu,\mu}
\]
where $\mu$ is defined in terms of $\mu_{I,\vertex}$ as in
(\ref{restricted}).  This agrees with the isotropic flux operator
defined in isotropic models,
\begin{equation}
 \hat{p}\sigma(|\mu\rangle)=\sigma(\hat{p}|\mu\rangle)\,,
\end{equation}
and shows in particular that $\hat{p}$, unlike $\hat{p}_{I,\vertex}$,
maps an isotropic distribution to another such state.  Thus, the
isotropic representation in loop quantum cosmology follows from the
inhomogeneous one along the lines of a symmetry reduction at the
quantum level. Notice that this leads directly to an operator for $p$
rather than $\tilde{p}$ without explicitly introducing the cell size
$V_0$. We just extend fluxes over the whole lattice in order to make
them homogeneous, such that the combination $V_0^{2/3}\tilde{p}=p$
automatically arises.

\subsection{Gravitons as collective excitations}

On inhomogeneous lattice states we can look for representations
of inhomogeneous excitations of geometry. Classically, such
excitations are given by modes of a metric perturbation
$q_{ab}=\bar{q}_{ab}+h_{ab}$ on a background $\bar{q}_{ab}$. In a
perturbative quantization on that background, $h_{ab}$ would thus be
the basic field to be quantized, leading to the expectation of
one-particle states called gravitons corresponding to
transverse-traceless modes of $h_{ab}$.

In loop quantum gravity, the situation is different because its basic
formulation is background independent. Having formulated here a
lattice description as a sector which re-introduces a metric
background, we can see how such metric perturbations are
realized. Simplifications due to Abelianization used here also occur
in constructions of explicit states in linearized gravity
\cite{Graviton}.  As in the full theory, basic excitations of the loop
quantization are given by dynamical moves which change the labels
$\mu_{I,\vertex}$ locally at a vertex to $\mu_{I,\vertex}\pm 1$. This
changes the local labels, corresponding to inhomogeneous modes, but
also the total volume corresponding to the background volume of
$\bar{q}_{ab}$. A quantum analog of the mode can be defined as the
difference
\[
 \delta p_{I,\vertex}=p_{I,\vertex}-\bar{p}=
 \frac{1}{2}\gamma\ell_{\rm P}^2\mu_{I,\vertex}-
(V(\{\mu_{I,\vertex}\})/N)^{2/3}
\]
between local labels and the average volume which captures properties of
the classical modes even at the level of effective dynamics
\cite{InhomEvolve}.

The quantum picture of these excitation is then very different from
the classical one: $\delta p_{I,\vertex}$ arises non-locally since we
have to sum over the whole lattice to subtract off the
background. Basic dynamical moves of the quantum theory change the
local quantum excitation as well as the background, leading to a mode
change also in a non-local way. In this picture, metric modes such as
gravitons arise as collective excitations out of the underlying
lattice formulation: many basic quantum modes combine to give
classical excitations.

\subsection{Quantum gravity corrections for large universes}

Quantum expressions differing from their classical analogs give rise
to corrections to classical equations. This usually occurs on small
length or large curvature scales as they are realized in the very
early universe. But subdivision in a lattice also makes relevant local
length scales smaller, and indeed we already noticed that inverse
power corrections as functions of $p^I(\vertex)$ become dominant over
holonomy corrections because the flux values are reduced on lattice
links if the lattice becomes finer. This opens the possibility for
small corrections on single lattice sites to add up to sizeable
corrections on the whole lattice, which can influence the evolution of
modes \cite{InhomEvolve} or even of the whole universe.

\section{Conclusions}

Models of loop quantum gravity are being investigated actively
regarding their phenomenological properties, and perturbative
inhomogeneities are currently being included. This brings us closer to
reliable computations of potentially observable properties such as
those of structure formation. It is then important to check all
intrinsic details of such models and see how faithfully they
incorporate features of the full theory. As we discussed, qualitative
effects are realized in homogeneous as well as inhomogeneous lattice
models in the same way. We introduced inhomogeneous lattice
constructions in a way which allows for a relation to a homogeneous
background. The relation to isotropic models is then clear, which
provides a new step toward relating isotropic models to the full
theory.  Although the lattice construction is analogous to that of
isotropic models, quantitative aspects can change which has a bearing
on which ranges of parameters one considers as natural or
unnatural. It also plays a role for which correction terms will be
dominant in different regimes which is the most important aspect for
phenomenological investigations.

The lattices used here are not intended as fundamental descriptions
since they are not background independent. They are rather to be
thought of as effective lattice models which result if some degrees of
freedom of the full theory are interpreted as providing a background
for other degrees of freedom. This simplifies the more irregular
fundamental behavior. It is this feature which helps to illustrate and
clarify some puzzling properties realized in isotropic models. There
are also further applications such as explicit constructions of
quantum field theories on a quantum spacetime as suggested in
\cite{QFTonCSTI,QFTonCSTII}, more general models for field propagation
with quantum corrections based on \cite{CorrectScalar}, or
cosmological perturbations \cite{InhomEvolve}.  Such lattice models
are more accessible for a relation to the full theory and can thus be
used to suggest relevant properties which one wants to ensure in
future constructions and specifications of the full theory at a basic
level.

For now, the construction already presents several applications. They
all rely on the re-introduction of a background which is necessary to
interpret computational results.  We focused here on kinematical
aspects, but an inclusion of characteristic dynamical ones is in
progress. For a detailed description of all effects expected in fully
inhomogeneous settings, remaining difficulties are an explicit
treatment of subdivision and the inclusion of non-Abelian
properties. But this is not required explicitly for effective
perturbations where the lattice description discussed here is
sufficient.  It allows one to derive perturbation equations for metric
modes around the background, to be used for instance for cosmological
structure formation, or to understand the emergence of graviton states
conceptually.

{\em Note added:} After this paper had been submitted to the journal
General Relativity and Gravitation, the preprint gr-qc/0607100 by
Kristina Giesel and Thomas Thiemann was posted. Although the
conceptual setup is quite different from the approach followed here,
it is clear that there is a convergence of different ideas pursued
recently in loop quantum gravity. Especially in combination with the
results of Sec.~\ref{SymmRed}, there is now close contact between the
full theory (understood as a general framework to set up background
independent quantum kinematics and dynamics without symmetry
assumptions) and loop quantum cosmology.

\section*{Acknowledgements}

The author thanks Abhay Ashtekar, Mikhail Kagan, Thomasz Pawlowski,
Parampreet Singh and Kevin Vandersloot for discussions.

\end{document}